# POSSIBLE MANIFESTATION OF TOPOLOGICAL SUPERCONDUCTIVITY AND MAJORANA BOUND STATES IN THE MICROWAVE RESPONSE OF THIN FeSe$_{1-x}$Te$_x$ FILM

N. T. Cherpak[1], A. A. Barannik[1], Y.-S. He[2], L. Sun[2], Y. Wu[2], S. I. Melnyk[1]

[1]O. Ya. Usikov Institute for Radiophysics and Electronics of National Academy of Sciences of Ukraine, Kharkiv, Ukraine
[2]Institute of Physics of Chinese Academy of Sciences, Beijing, PRC

ABSTRACT

The paper analyzes the characteristics of the microwave (MW) response of FeSe$_{1-x}$Te$_x$ films based on the results of measuring the impedance properties of the films in the X-band for two orientations of the film in the MW magnetic field, perpendicular and parallel.
The analysis of the temperature dependence of the microwave response of a film with a perpendicular orientation (in which the peculiarity of the response is manifested) was carried out by means of physical considerations, taking into account also the results of the research of this superconductor by other authors using the ARPES technique and tunneling spectroscopy.
It was concluded that with perpendicular orientation, two competing mechanisms of MW energy dissipation in the film can occur, one of which leads to the increase in energy dissipation caused by magnetic vortices with an MW field, and the other to its decrease due to the emergence of Majorana bound states with zero energy

## 1. INTRODUCTION

More than ten years have passed since the discovery of a strange feature of the response of a thin film of the then new superconductor FeSe$_{1-x}$Te$_x$ in a perpendicular ($\perp$) microwave (MW) field [1,2]. The peculiarity of the response is experimentally manifested in the form of peak non-monotonicity with an extremum at $T_m < T_c$ in the temperature dependence of the Q-factor of the resonator with the film. The response is strange because it is not accompanied by any features in the temperature dependence of the frequency of the resonator with the film, that is, in the temperature dependence of the surface reactance in this orientation. In addition, it is completely absent in the parallel ($\parallel$) orientation of the film [3,4]. And, what was most surprising, this feature was manifested only in the films of FeSe1-xTex superconductors and was not manifested in the films of other type-II superconductors [5].

Thus, the question arises about the physical properties of the FeSe$_{1-x}$Te$_x$ superconductor, which distinguish it from the general family of superconductors-II, including iron-containing ones (Fe-superconductors).

Recently, for 10-15 years, the hot topic of manifestation of topological features in Fe-superconductors has been actively developing in superconductivity physics (see, for example, a review [6]). The fundamental concept, foundations of the theory, expected properties and implementation of topological superconductors (TSC) with a deep understanding of the problem are outlined in [7]. Ideas for the implementation of TSC appeared only after the discovery of topological insulators [8]. Interest in TSC is closely related to Majorana fermions, which are simultaneously their own antiparticles [9]. In the FeSe$_{1-x}$Te$_x$ superconductor, using the ARPES technique, surface states with the dependence of the electron energy on the momentum in the form of a Dirac cone (as in graphene) were detected [10]. At the same time, the bulk properties of the superconductor remain. This became the prerequisite for the appearance of Majorana fermions in this superconductor. It has been shown theoretically that they can create stable so-called Majorana bound states (MBS) on the surface of a superconductor in regions where magnetic vortices arise [11].

A number of works have been performed in which the appearance of a clear signal at zero bias voltage was demonstrated using the technique of scanning tunneling spectroscopy [6,10,12]. This result was enthusiastically interpreted as the detection of a zero-energy MBS. The latter is also called Majorana zero mode (MZM) because it is characterized by zero energy. Over time, a certain skepticism appeared, related to the fact that there may be other physical phenomena with an energy spectrum, in which the lowest energy level can be close to the zero level. There was a question about the final identification of MZM in tunnel experiments and the desirability of other approaches to such identification [13,14].

Therefore, in the $FeSe_{1-x}Te_x$ studies, an intriguing situation arose, when, on the one hand, unexplained peculiarities of the response of thin films in the MW magnetic field were revealed and, on the other hand, the MW research technique may become critical for the final identification of bound Majorana states in these superconductors. The mentioned aspects stimulated the analysis of the experimental results of the study of $FeSe_{1-x}Te_x$ films from the point of view of a possible connection between the detected MX features of the film response and the topological features of this superconductor. Such an approach makes it possible to formulate non-trivial problems in the field of microwave electrodynamics of magnetic vortices in thin superconducting films, including films of topological superconductors with Majorana bound states, depending on temperature.

## 2. RESPONSE OF $FeSe_{1-x}Te_x$ FILM AS A FUNCTION OF TEMPERATURE AND FILM ORIENTATION IN MW MAGNETIC FIELD

Experimentally, the MW response of the films was studied in a cylindrical resonator with copper walls with $H_{011}$ mode in the X-band based on the measurement of Q-factor and the eigen frequency of the resonator with and without the film in the temperature range from about 1.5K to the critical temperature $T_c$ and higher. The temperature of the resonator walls was kept constant at 4.2K. Two $FeSe_{1-x}Te_x$ films were studied, first with $x=0.7$ for the perpendicular ($\perp$) orientation of the film in the MW magnetic field $\mathbf{H}_\omega$, [1,2], then for $\perp$ and parallel ($\parallel$) orientations [3,4]. The thickness of the films measuring ~1x1mm$^2$ was equal to 100 nm. The substrates were single-crystal $LaAlO_3$ in the first case ($x=0.7$) and $CaF_2$ in the second case ($x=0.5$) with a thickness of 0.5 mm in both cases.

Fig. 1 presents the temperature dependences of Q-factor ($Q_\perp$) and eigen frequency $f_\perp$ for the $FeSe_{1-x}Te_x$ film ($x=0.7$), from which it is immediately clear that a certain decrease of $Q_\perp$ is observed, that is, an increase in the dissipation of MW energy in the region of a certain temperature $T_m < T_c$ ($T_c$=14.8 K) followed by some increase in this Q-factor upon reaching the critical temperature

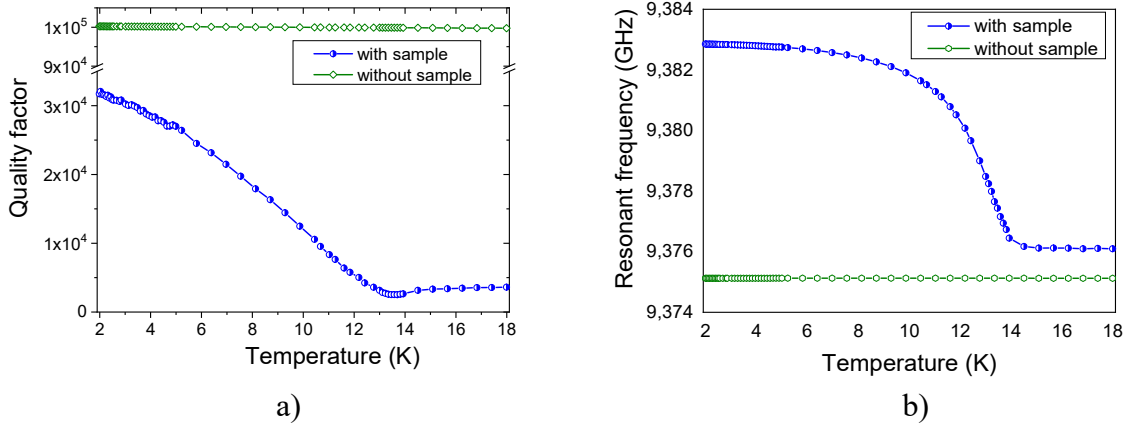

Fig. 1. Temperature dependence of $Q_\perp$ (a) and frequency $f_\perp$ (b) of the resonator with FeSe$_{1-x}$Te$_x$ ($x$=0.7) film at the $\perp$ orientation of the film in the MW magnetic field of the resonator. The temperature dependences of the Q-factor (a) and the frequency f (b) of the resonator without a film, but with a dielectric substrate, are also presented.

The next experiment was performed with the FeSe$_{1-x}$Te$_x$ film ($x$=0.5) for the $\perp$ and $\parallel$ orientations of the film [3, 4]. It showed that, with the $\perp$ orientation of the film, the feature of the MW response is preserved, despite the change in the concentration of Te ions, and the critical temperature ($T_c$ =18.8K). The change in the latter is obviously caused by a change in the concentration $x$ of Te ions [15]. Note that the temperature $T_m$ (with the peak maximum value of the inverse Q-factor) has become closer to $T_c$. At the same time, any non-monotonicity in the temperature dependence of the Q-factor of the resonator with the film for the $\parallel$ orientation was not observed (Fig. 2). It was also absent in the frequency dependence of the resonator with the film for the same orientation.

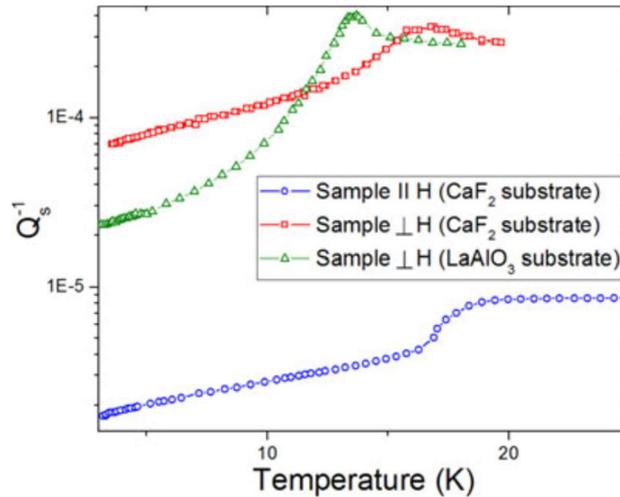

Fig. 2. Temperature dependence of the inverse Q-factor of the resonator (decrement) with FeSe$_{1-x}$Te$_x$ film for the $\perp$ orientation ($x$ = 0.5 and 0.7) and for the $\parallel$- orientation ($x$ = 0.5).

Experiments were also performed with a number of new FeSe$_{1-x}$Te$_x$ film ($x\cong0.5$), that do not exhibit the specified features of the MW response in the $\perp$ orientation. There are two possible reasons for the disappearance of this feature in the latest copies of the film. One of the reasons may be the use of a new resonator for measuring the MW response. In it, unlike the resonator used above, the body is made of low-temperature superconductor Nb, while the Q-factor has become an order of magnitude higher. This means a significant increase in the amplitude of the MW magnetic field in the resonator with a possible effect on the features of the interaction of the field with the superconducting film in the ⊥ direction. In addition, measurements were carried out at frequencies 2-3 times higher than the frequencies in works [1-4]. The second reason may be related to the subtle dependence of the topological features of the films on the chemical composition of the films [16]. It is even proposed to write the composition of this superconductor as $Fe_{1+y}Se_{1-x}Te_x$ to emphasize the influence of $x$ and $y$ on the properties of the superconductor.

## 3. PHENOMENOLOGICAL ANALYSIS OF THE FEATURES OF MW RESPONSE OF FeSe$_{1-x}$Te$_x$ FILMS AND THE POSSIBLE MANIFESTATION OF BOUND MAJORANA STATES IN THE PERPENDICULAR ORIENTATION OF THE FILM IN MW MAGNETIC FIELD

In this section, the goal is to analyze the features of the MW response of the FeSe$_{1-x}$Te$_x$ film, making maximum use of the experimental results of measuring the response of the H$_{011}$ mode resonator with the film (see section 2), as well as phenomenological considerations based on the results of the study of this unique superconductor by other methods involving other experimental techniques. An important factor for the authors was also the comparison of the MW response of FeSe$_{1-x}$Te$_x$ films with the response of the films of other superconductors and their thick samples (including FeSe$_{1-x}$Te$_x$ single crystals).

Next, it is more convenient to speak about the dissipation of MW energy in a resonator with a film, so we will move on to the value of the effective surface resistance $R_{S\perp}{}^{eff}$ of the film with the ⊥ orientation ~ $Q_{s\perp}^{-1}$, which depends on the film thickness and is used in the study of the impedance properties of superconductor films (see, for example, [2]).

In fig. 3. the experimental temperature dependence of $R_{S\perp}{}^{eff}$ for FeSe$_{1-x}$Te$_x$ films with $x=0.7$ is given, together with the estimated dependences $\tilde{R}_{S\perp}^{eff}(T)$ and $R_{S\perp 0}^{eff}(T)$. The last two, respectively, show the expected temperature dependence of the effective surface resistance in the presence of magnetic vortices in the film, taking into account the demagnetization factor (by analogy for the films of other known superconductors-II), and the dependence of the surface resistance in the apparent absence of magnetic vortices in the film (by analogy with the orientation of the film with parallel orientation).

If we look at the temperature dependence of $R_{S\perp}{}^{eff}$, we can immediately assume that when $T$ increases to $T_c$ in the superconductor, two competing physical mechanisms operate, one of which causes an increase in the dissipation of MW energy, and the other, on the contrary, leads to its decrease. At the temperature $T_m < T_c$, the maximum dissipation occurs, after which the energy dissipation decreases even in some interval at $T > T_c$, which contrasts with the behavior of films of other superconductors. It is possible to further imagine that when the energy dissipation reduction mechanism is turned off while the dissipation growth mechanism is preserved (obviously due to the vortex structure), the effective resistance would have a temperature dependence similar to the curve $\tilde{R}_{S\perp}^{eff}(T)$ in Fig. 3. At the same time, it is clear that $\tilde{R}_{S\perp}^{eff}(T) = R_{S\perp 0}^{eff}(T)$ at $T = T_c$ because magnetic vortices should be absent in the normal state of a superconductor-II with known properties.

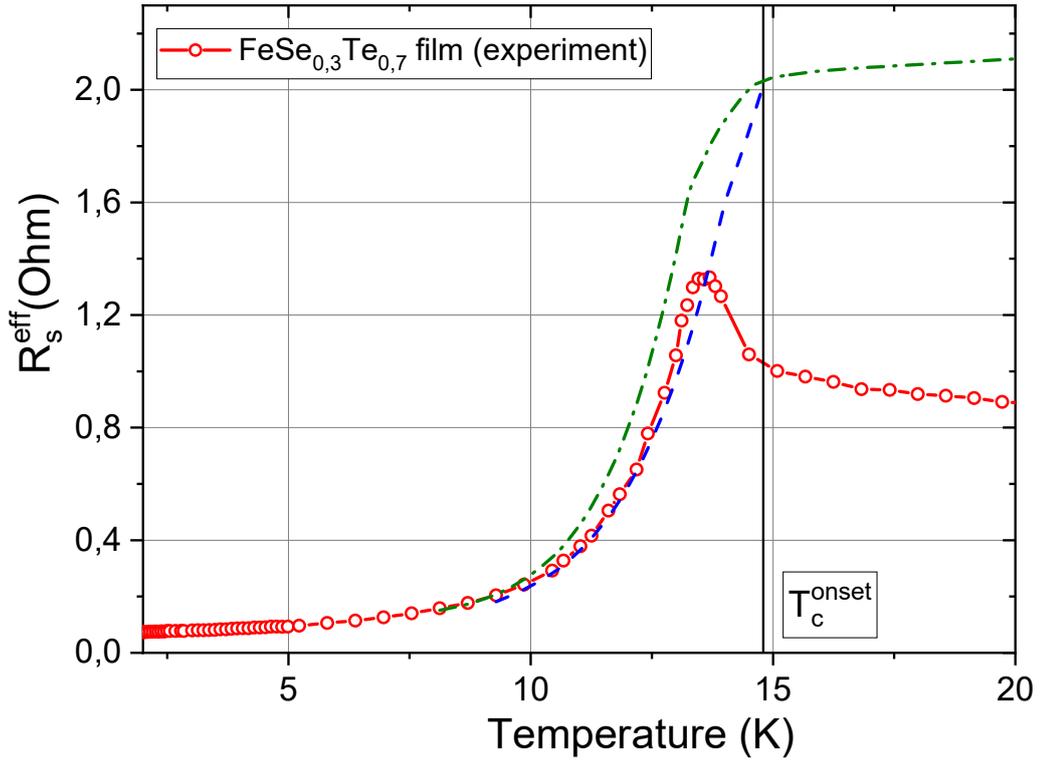

Fig. 3. Temperature dependence of the effective surface resistance of the $FeSe_{1-x}Te_x$ film with $\perp$ orientation; o – $R_{S\perp}^{eff}(T)$, dashed line – $R_{S\perp 0}^{eff}(T)$, dash-dotted line – $\tilde{R}_{S\perp}^{eff}(T)$; see details in the text.

A fundamental question arises in this model. Why $R_{S\perp 0}^{eff}(T) > R_{S\perp}^{eff}(T)$ at $T \geq T_c$. It can be imagined that the reason is the same circumstance that leads to a decrease in $R_{S\perp}^{eff}(T)$ even when $T > T_c$. This forces us to make the important assumption, namely, in the $FeSe_{1-x}Te_x$ thin film at the $\perp$ orientation of the MW magnetic field (under the action of both dissipation factors) has the features that are completely absent in the film with the $\parallel$ orientation.

Thus, based on the results of measuring the MW response of thin films at the $\perp$ and $\parallel$ orientations in the MW magnetic field and comparing these results with data obtained by other researchers using ARPES and tunneling spectroscopy techniques, the authors hypothesize a cardinal influence of topological features and Majorana bound states on the MW properties of $FeSe_{1-x}Te_x$ superconductor films. In this assumption, the main role should be played by MBSs, which in the proposed scenario form a mechanism for reducing the dissipation of MW energy in the superconductor film. The assumption stimulates the formulatjon of a number of fundamental problems in the physics of superconductivity and the dynamics of quantum systems. The solutions should provide answers to the nature of the identified features of the MW response of topological superconductor films, in particular, to the nature of the interaction of the MW magnetic field with MBS in composition with the magnetic vortices.